\documentclass[reprint,
amsmath,amssymb,
aps
]{revtex4-2}
\usepackage{graphicx}
\usepackage{dcolumn}
\usepackage{bm}
\usepackage{amsthm}
\usepackage{mathrsfs}
\usepackage{xcolor}
\usepackage{textcomp}

\usepackage[colorlinks=true, linkcolor=blue, citecolor=blue, urlcolor=blue]{hyperref} 

\begin{document}
\title{Cascade Crack in Chain of Beads}
\author{Meysam Bagheri}
\email{meysam.bagheri@fau.de}
\author{Thorsten P\"oschel}
\email{thorsten.poeschel@fau.de}
\affiliation{Institute for Multiscale Simulation, Friedrich-Alexander-Universität Erlangen-Nürnberg}

\date{\today}

\begin{abstract}
We consider a homogeneous chain of spheres linked by liquid bridges under tension. The rupture of a single liquid bridge leads to a fragmentation cascade driven by the inverse relation between the capillary force and the sphere distances. The initial length of the liquid bridges determines the number and size of the fragments and the velocity of the fragmentation front.  
\end{abstract}

\maketitle
\section{Introduction}
The nucleation and propagation of cracks in materials are fundamental concerns in several areas of science and engineering \cite{goehring2015desiccation,schneider2017,hao2019atomistic,cochard2024propagation, breakage2025particle}. Understanding how cracks propagate and navigate their path within diverse materials – from brittle solids to drying colloidal suspensions – is crucial for predicting material failure and optimizing design \cite{dufresne2006dynamics,dugyala2016role,gao2017theoretical,zhao2022examination}. While the specific mechanisms of crack propagation differ between solids and drying suspensions, crack initiation often occurs at inherent weaknesses within the material. External stress can nucleate a crack from a preexisting flaw in brittle solids, such as a microcrack or impurity. In drying, shrinkage or capillary forces often induce internal stresses, which can initiate a crack at a void or a weak sphere-sphere bond in drying colloidal suspensions \cite{mondal2023physics}. In either case, the inception of the first crack or bond breaking results in stress relaxation \cite{skjeltorp1988fracture,audoly2005fragmentation,routh2013drying,goehring2015desiccation,bourrianne2021crack}. This stress relaxation may induce strain and additional stress concentration, initiating a cascade of subsequent breaking events. A remarkable example of this phenomenon in solids is the \textit{spaghetti problem} that perplexed Richard Feynman. He observed that a bent spaghetti noodle does not break in half but rather into multiple pieces. Audoly and Neukrich \cite{audoly2005fragmentation} later revealed the underlying physics: the initial break at the point of highest curvature triggers a sudden stress relaxation, generating powerful flexural waves that cascade down the rod, causing further breaks \cite{audoly2005fragmentation}. Inspired by spaghetti's similarity to elastic rods,  a ubiquitous element in nature and engineering, studies investigate how twisting can control fracture cascades and reduce the number of fragments \cite{heisser2018controlling}, and the correlation between a rod's aspect ratio and the number of pieces it breaks into \cite{zhang2022number}.

We consider a related phenomenon in a chain of beads connected by liquid bridges, which can be regarded as a one-dimensional colloidal suspension in the final stage of drying \cite{Zhou2006, routh2013drying, goehring2015desiccation, rozynek2022fabrication}. One-dimensional beaded chains have also recently found applications in fabricating new materials \cite{rozynek2022fabrication, dutka2017continuous}. Here, liquid bridges between the beads exert attractive forces, establishing chains. The rupture of one of the bridges, typically due to the evaporation of the liquid \cite{goehring2015desiccation}, leads to a perturbation front followed by a fragmentation cascade of cracks of other bonds. 

\section{Model}

We consider a homogeneous chain of length $L$ composed of $n$ spheres of radius $R$ connected by $n-1$  liquid bridges. The centers of the first and last spheres are fixed at $x_1=0$ and $x_n=\left(n-1)(2R+S^{(0)}\right)$, respectively, where $S^{(0)}$ is their initial value of the distance between the surfaces of adjacent spheres,
$S_i\equiv x_{i+1}-x_i-2R$, $i\in\{1,n-1\}$. The value $S_i>0$ describes the length of the corresponding liquid bridge. If $S_i\le 0$, the spheres are in contact, and $-S_i$ denotes their mutual deformation. The force between adjacent spheres, $i$ and $i+1$, reads
\begin{equation}
F_i (S_i) = \begin{cases}
  F^\text{cap}_i(S_i), & \text{if } S_i> 0 \\
  F^\text{cap}_i(0) + F_i^\text{n}(S_i), & \text{if } S_i\le 0\,,
\end{cases}
\label{totalforce}
\end{equation}
where $F^\text{cap}_i$ stands for the capillary force due to the liquid bridge and $F_i^\text{n}$ is the contact force between the spheres.

The capillary force results from the solution of the Young-Laplace equation and depends on the length, $S_i$, and the volume, $V$, of the enclosed liquid. The coefficient of surface tension, $\gamma$, and the contact angle, $\theta$, between the sphere surface and the surface of the liquid bridge enter as material constants. The Young-Laplace equation cannot be solved in closed form, but several highly precise fit formulae of the numerical solution are available in the literature. 
Here, we chose the approximation in \cite{bagheri2024approximate}. The force between viscoelastic spheres in contact is \cite{hertz1881beruhrung,brilliantov1996model}
\begin{equation}
    F^\text{n}_i = \min\left(0, -\rho (-S_i)^{3/2} - \frac{3}{2}A\rho\sqrt{(-S_i)}{(-\dot{S}_i)}\right)
\end{equation}
where  
\begin{equation}
    \rho \equiv \frac{2ER}{3(1-\nu^2)} 
    \label{eq:Hertz_rho}
\end{equation}
involves the elastic modulus, $E$, and the Poisson ratio, $\nu$. \autoref{tab:material_parameters} gives the material and system parameters for polystyrene microspheres connected by water liquid bridges, which are often
used in studies of colloidal self-assembly \cite{Zhou2006, kolegov2019joint}. 

With these parameters and assuming the coefficient of restitution 0.3 for spheres colliding at $1$ m/s, we obtain $A=2.53\times{10}^{-9}$s \cite{muller2011collision}.
\begin{table}[htbp]
\caption{Simulation parameters.}
\label{tab:material_parameters}
\begin{center}
\begin{tabular}{l@{\quad}l@{\quad}ll}
\hline
variable & value & unit \\[2pt]
\hline
sphere radius ($R$) & 1  & $\mu\text{m}$\\
sphere density & 1050 & kg/m$^3$\\
elastic modulus ($E$)  & 3.2 & GPa\\
Poisson ratio ($\nu$)  & 0.3 & \\[2pt]
contact angle ($\theta$)  & 27.5 & degree\\
surface tension & 0.0728 & N/m\\
liquid bridge volume ($V$)  & $10^{8}$ & n$\text{m}^3$\\
\hline                
\end{tabular}
\end{center}
\end{table}

For any value of $S_i<0$, the repulsive Hertz contact force is larger than the attractive capillary force by orders of magnitude. Therefore, for spheres in contact, in \autoref{totalforce} we approximate the capillary force by its contact value, $F^\text{cap}_i(S_i)\approx F^\text{cap}_i(0)$. The liquid bridge volume considered here is sufficiently large such that surface forces, e.g., the van der Waals force, are negligible compared to the capillary force between $\mu$m-sized spheres \cite{Yang2016}.

The solution of the Young-Laplace equation is linearly stable for any length, $S_i$, of the liquid  ridge. In practice, it breaks when the surface energy required to cut the bridge at its narrowest point approaches the value of the thermal fluctuations. For many practical purposes, the empirical formula \cite{Willett2000,bagheri2025discrete}
\begin{equation}
    S^\text{r} = \frac{1}{R}\left( 1+\frac{\theta}{2}\right) \left( V^{1/3} +\frac{1}{10 R} V^{2/3}\right)
    \label{eqScrit}
\end{equation}
is appropriate to describe the maximal elongation. Any further elongation would result in rupture. In the following, we will use the parameter 
\begin{equation}
    \alpha \equiv \frac{S^\text{r}}{S^{(0)}}\,,\qquad  1.01\le \alpha\le 4.5
    \label{eq:alphadef}
\end{equation}
where $1.01$ corresponds to a narrow liquid bridge close to rupture. As $\alpha$ increases, the liquid bridge becomes shorter and thicker.

We will show that the dynamics of the spheres driven by capillary force imbalances leads to characteristic fragmentation patterns. In experiments with thin liquid films containing particle monolayers \cite{Zhou2006,li2017dewetting, kolegov2019joint} it was shown that this instability leads to dry patches of particles within the monolayer. 

\section{Results}

We consider a homogeneous chain of spheres at equilibrium where $S_i=S^{(0)}$ and all forces are balanced. Note that this equilibrium is unstable, thus, any fluctuation, $\left|S_i-S^{(0)}\right|\ne 0$, would grow. We will come back to this property below.  

\begin{figure}[htbp]
\includegraphics[width=\columnwidth]{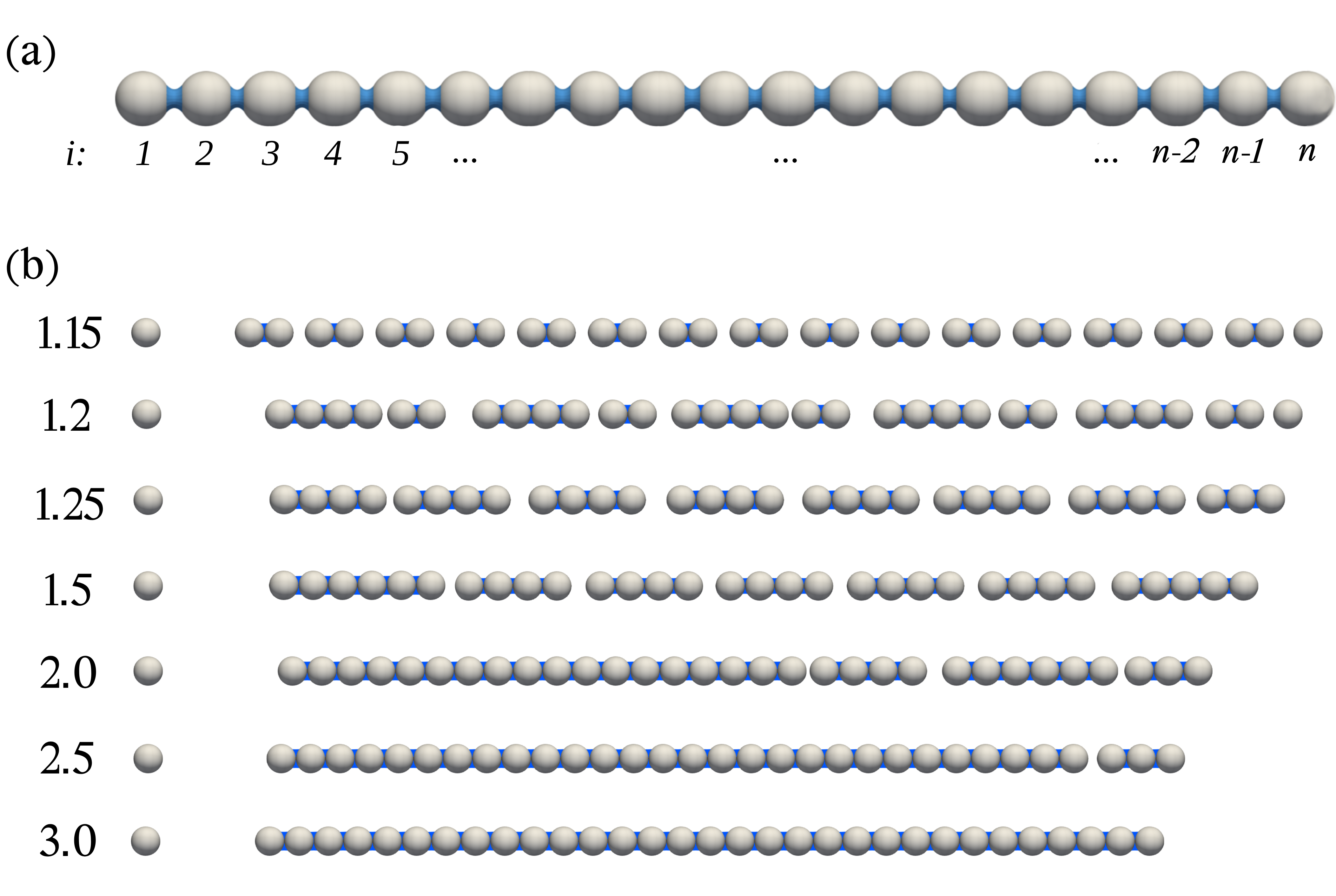}
\caption{\label{FigPhenomenon} (a) Initial configuration of the chain. (b) Final configurations obtained for a chain of $n=32$ spheres for $\alpha \in\left\{1.15,1.2,1.25,1.5,2,2.5,3\right\}$}
\end{figure}

At time $t=0$, we cut the liquid bridge between spheres 1 and 2 (see \autoref{FigPhenomenon}(a)), which perturbs the second sphere's initial local equilibrium, ($F^{\text{cap}}_1=F^{\text{cap}}_2$), and causes its accelerated motion toward the third sphere. This, in turn, causes $F^{\text{cap}}_2> F^{\text{cap}}_3$ (since $F^{\text{cap}}_i\propto S^{-1}_i$), pulling the third sphere towards the second one. 

The initiated perturbation propagates through the chain, inducing subsequent capillary force imbalances. As a consequence, all spheres, $i=2,\dots n-1$, will move collectively in positive $x$-direction. The individual motion of each sphere results from a complicated interplay of the non-linear force due to the liquid bridges and the particle's inertia. Movie S1 of the supplemental material shows an example of the chain's dynamics. The interesting phenomenology arises from the fact that a liquid bridge breaks as soon as it reaches its maximum expansion, $S^\text{r}$, given by \autoref{eqScrit}. The simulation terminates when all intact bridges reach a length of zero and no further breaks can be achieved. The resulting fragments are thus subchains whose spheres are connected by intact liquid bridges. \autoref{FigPhenomenon}(b) shows the final configurations for a chain of $n=32$ spheres for various values of $\alpha$. 

For small $\alpha \lesssim 1.15$, the liquid bridges are initially highly elongated, close to the critical value $S^\text{r}$. As soon as the first bridge between particles 1 and 2 is cut at time $t=0$, sphere 2 moves towards sphere 3. The reduced distance $S_2$ leads to an increase in force $F_2$. As a result, particle 3 is no longer in equilibrium, since $\left|F_2\right| > \left|F_3\right|$. Consequently, sphere 3 also moves towards ball 2. This motion in turn increases the distance $S_3$, and so on. The described sequence can be seen in Movie S1 of the supplemental material. 

The position $x^\text{per}$ of the propagating perturbation can be identified by the force imbalance: The front arrived at the bridge between spheres $i$ and $i+1$ when  
%\begin{equation}
%\begin{split}
%{\left|{F^{\text{cap}}_{i}(t=t^{\text{per}}_{i}) - F^{\text{cap}}_{i}(t=0)}\right|} & > \varepsilon\\
%{\left|{F^{\text{cap}}_{i+1}(t=t^{\text{per}}_{i}) - F^{\text{cap}}_{i+1}(t=0)}\right|} &\le \varepsilon\,.
%\end{split}
%\end{equation}
\begin{equation}
\begin{split}
\left|\frac{ F^{\text{cap}}_{i}(t=t^{\text{per}}_{i})}{F^{\text{cap}}_{i}(t=0)}\right|-1 & > \varepsilon\\
\left|\frac{F^{\text{cap}}_{i+1}(t=t^{\text{per}}_{i})}{F^{\text{cap}}_{i+1}(t=0)}\right|-1 &\le \varepsilon\,.
\end{split}
\end{equation}
(In the simulation we chose $\varepsilon = 10^{-16}$.) The position of the propagating front at this time is then
\begin{equation}
    x^{\text{per}} = \frac{{x}_{i}\left(t^{\text{per}}_{i}\right) + {x}_{i+1}\left(t^{\text{per}}_{i}\right)}{2}\,.
    \label{eq:vel perturbation}
\end{equation}

With propagating front, the total mass of particles with $x_i<x^\text{per}$ increases such that this part of the chain cannot follow the front under the action of the pulling liquid bridge. Consequently, with propagating front, the lengths of the liquid bridges increase until they exceed the value $S^\text{r}$, at which the bridge breaks and a fragment is formed. {From this point on, the process repeats itself in exactly the same way as after the first bridge between spheres 1 and 2 was cut. From this argument, it immediately follows that the fragments are approximately the same length, apart from fluctuations.}

\begin{figure}[htbp]
\includegraphics[width=9cm]{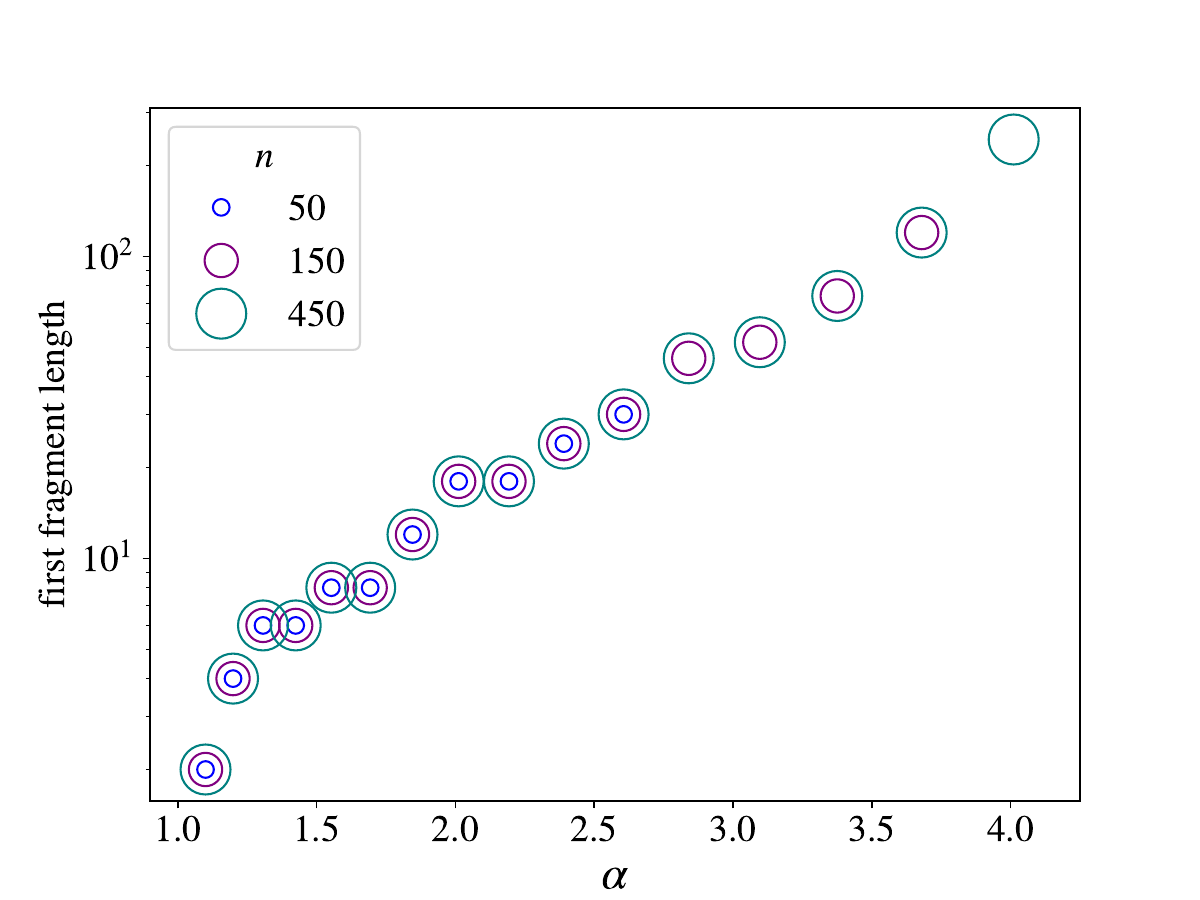}
\caption{\label{FigInitialFragment}
The length of the first fragment (number of spheres) is nearly an exponential function of $\alpha$, independent of the chain size $n$.
% \\ \TP{replace axis label of vertical axis: ``first fragment length''}
}
\end{figure}

With increasing $\alpha$ (shorter initial lengths of the bridges), a longer section of the chain is required until the fragmentation length is reached according to the scenario described above, see  \autoref{FigPhenomenon}(b). \autoref{FigInitialFragment} shows the length of the first fragment as a function of $\alpha$. 
The size of the fragments increases nearly exponentially with $\alpha$, independent of the chain size.

\begin{figure}[htbp]
\includegraphics[width=9cm]{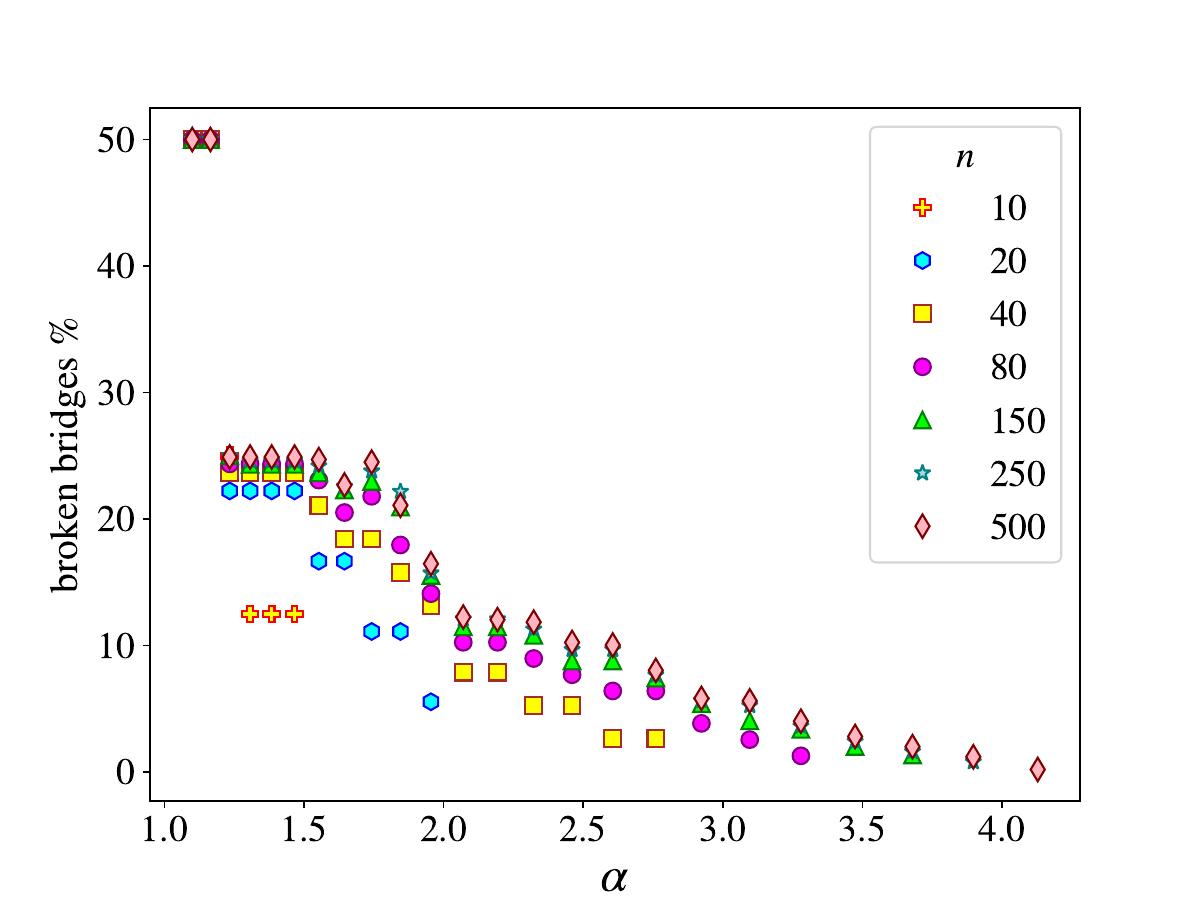}
\caption{\label{fragments-alpha}
Fraction of broken bridges as a function of $\alpha$ for several chain lengths, $n$. 
}
\end{figure}

\autoref{fragments-alpha} shows the proportion of ruptured liquid bridges at the end of the process as a function of $\alpha$ for various chain lengths $n$.
For $\alpha \lesssim 1.15$, approximately 50$\%$ of bridges break, regardless of the chain length, indicating dimer formation. The percentage decreases with increasing $\alpha$ in agreement with increasing fragment size. Zhang et al. \cite{zhang2022number} observed a similar trend in their study on the \textit{spaghetti problem}, where an increased diameter-to-length ratio of the spaghetti strands led to fewer fragments. Since the sphere radius remains constant in our simulations, the parameter $\alpha$ in our study is proportional to the diameter-to-length ratio in \cite{zhang2022number}.

\begin{figure}[htbp]
\includegraphics[width=9cm]{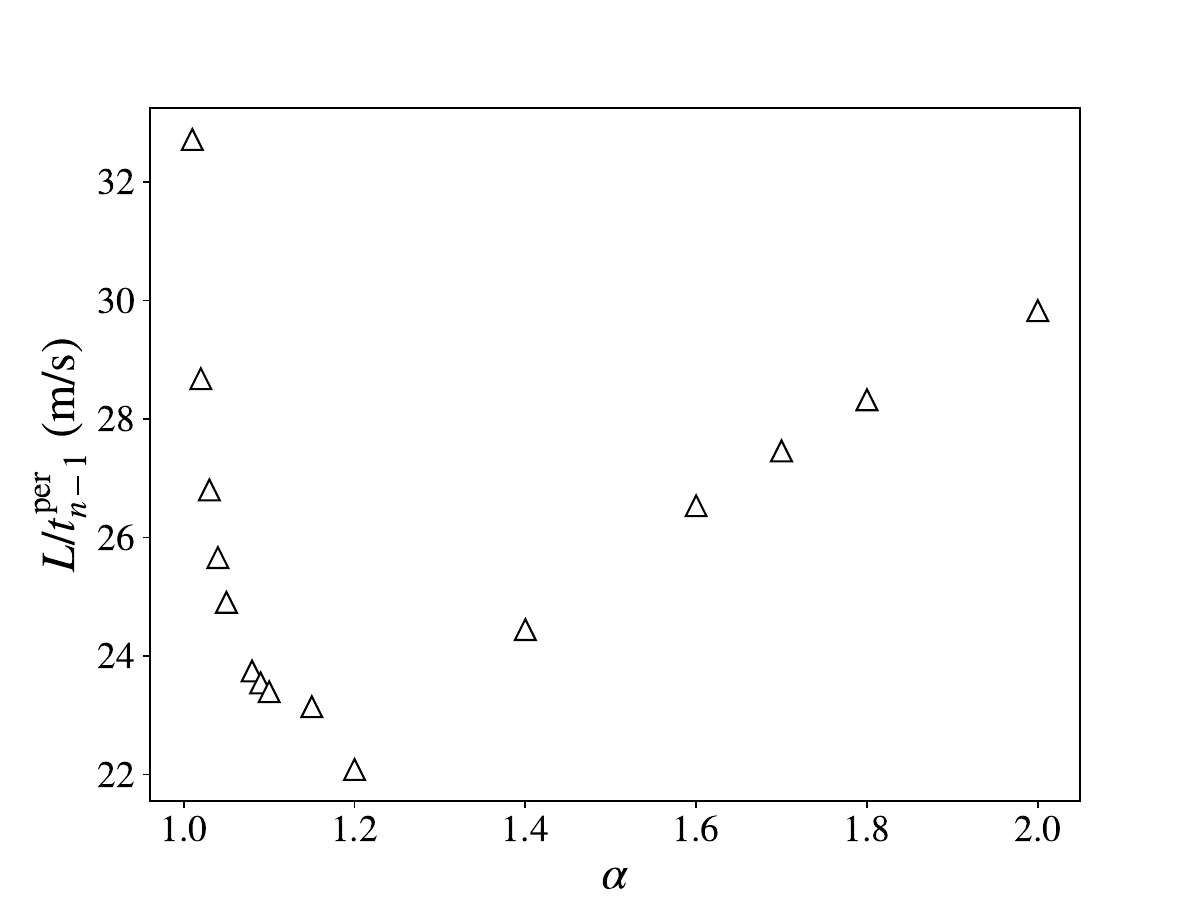}
\caption{Propagation velocity of the perturbation, $L/t^\text{per}_{n-1}$, as a function of $\alpha$.
\label{fig:extra}
% \TP{remove color}
}
\end{figure}

The initial extension of the liquid bridges affects not only the length of the fragments but also the propagation verlocity of the perturbation front. 
%\autoref{fig: Tper-L-F} 
\autoref{fig:extra} shows the velocity $L/t^{\text{per}}_{n-1}$ of the front for a chain of length $n=32$ as a function of $\alpha$. 
%\begin{figure}[htbp]
%\includegraphics[width=9cm]{figures/fig4-modified.pdf}
%\caption{\label{fig: Tper-L-F}
%$t^\text{per}_{n-1}$ as a function of $\alpha$ and the corresponding initial capillary force $F^{\text{cap}}$ of the liquid bridges for a chain with $n=32$.}
%\end{figure}

With decreasing $\alpha$ ($2.0 \rightarrow 1.12$), the initial capillary force becomes weaker, slowing down the dynamics of the spheres and resulting in slower propagation front. Upon reaching the dimer formation regime ($\alpha \lesssim 1.12$), see \autoref{fragments-alpha}, the velocity increases sharply. For $\alpha\lesssim 1.12$, although the capillary force is small, it overcompensates the reduced inertia of the short fragments. Moreover, for $\alpha\lesssim 1.12$ the liquid bridges are close to the rupture threshold $S^\text{r}$, thus a small perturbation leads to fragmentation, facilitating faster propagation of the perturbation front. 

\begin{figure}[htbp]
\includegraphics[width=\columnwidth]{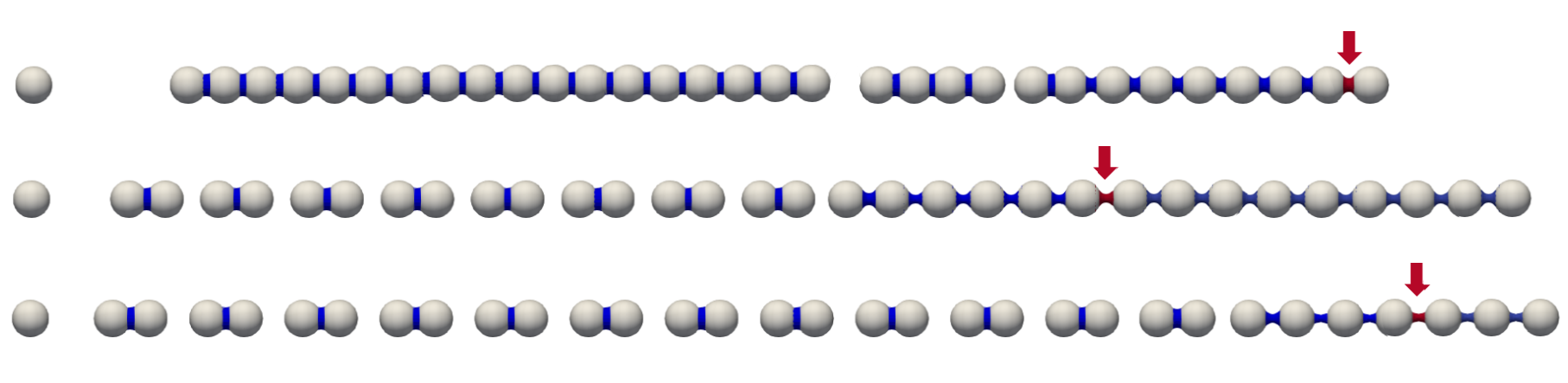}
\caption{\label{perturbationFront}
Snapshots of chains with $n=32$ and $\alpha\in\left\{2,1.1,1.01\right\}$ from top to bottom. Red arrows indicate the position of the perturbation front the same time instant}
\end{figure}

The non-monotoneous dependence of the propagation velocity on $\alpha$ and, thus, on the chain length $L=(n-1)\left(2R+S^\text{r}/\alpha\right)$ is shown in \autoref{perturbationFront}. 
The figure shows three chains of length $n=32$ with different $\alpha$ at the same instant of time. Red arrows mark the position of the perturbation front at this time.

\section{Conclusion}
We studied the propagation of a perturbation front in a chain of $n$ spheres connected by liquid bridges. At time $t=0$ the perturbation was initiated by cutting the bridge between the first spheres. The front propagates through the chain at a velocity that depends on the initial elongation. The interplay between the force perturbed balance of the liquid bridges and the spheres' inertia results in a cascade of ruptures of liquid bridges, generating fragments of a characteristic length. Both the fragment size and the velocity of the perturbation front depend on the initial extension of the liquid bridges. Note that the initial configuration of a chain of evenly spaced identical spheres with identical liquid bridges represents an unstable equilibrium. Any microscopic random perturbation which is always present in physical systems, would drive the system out of equilibrium. Therefore, our study is restricted to small values of $n$ so that the time required for the perturbation front to pass through the chain, $t^\text{per}_{n-1}$, is short compared to the time required for a microscopic random fluctuation to cause a noticeable displacement of any of the spheres. The validity of this condition has been verified for all simulation results presented here.

% \bibliography{cascade-library}

\end{document}